\documentclass[prl,twocolumn,superscriptaddress,showpacs]{revtex4}
\usepackage{times}
\usepackage{amsmath}
\usepackage{amsfonts}
\usepackage{graphicx}
\usepackage{pst-all}

\begin{document}

\title{Numerical Calculation of the Neutral Fermion Gap at $\nu=5/2$}
\author{Parsa Bonderson}
\affiliation{Microsoft Research, Station Q, Elings Hall, University of California, Santa Barbara, CA 93106, USA}
\author{Adrian E. Feiguin}
\affiliation{Department of Physics and Astronomy, University of Wyoming, Laramie, WY 82071, USA}
\author{Chetan Nayak}
\affiliation{Microsoft Research, Station Q, Elings Hall, University of California, Santa Barbara, CA 93106, USA}
\affiliation{Department of Physics, University of California, Santa Barbara, CA 93106, USA}
\date{\today}

\begin{abstract}
We present the first numerical computation of the
neutral fermion gap, $\Delta_\psi$,
in the $\nu=5/2$ quantum Hall state, which is analogous
to the energy gap for a Bogoliubov-de Gennes
quasiparticle in a superconductor. We find
$\Delta_\psi \approx 0.027 \frac{e^2}{\varepsilon \ell_0}$,
comparable to the charge gap, and
discuss the implications for topological quantum information processing.
We also deduce an effective Fermi velocity $v_F$ for neutral fermions
from the low-energy spectra for odd numbers of electrons, and
thereby obtain a correlation length
$\xi_{\psi}={v_F}/\Delta_{\psi} \approx 1.3\, \ell_0$.
We comment on the implications of our results for
electronic mechanisms of superconductivity more generally.
\end{abstract}

\maketitle









%

The $\nu=5/2$ fractional quantum Hall state~\cite{Willett87,Pan99b,Eisenstein02} has
been the subject of intense experimental and theoretical
investigation in recent years because it may
support non-Abelian anyons and may
serve as a platform for topological quantum information processing~\cite{Kitaev97,Freedman98,Nayak08}. Theoretical~\cite{Morf98,Rezayi00,Feiguin08,Peterson08,Feiguin09,Bishara09a,Rezayi09}
and experimental~\cite{Radu08,Dolev08,Willett09} evidence
has been rapidly accumulating in favor of $\nu=5/2$ being an Ising-type non-Abelian state,
in the universality class of either the Moore-Read (MR) Pfaffian
state~\cite{Moore91,Greiter92} or the anti-Pfaffian ($\overline{\text{Pf}}$) state~\cite{Lee07,Levin07}.

The potential use of this state for
topological quantum information processing is dependent on the size of
the energy gaps $\Delta_a$ to different species of
quasiparticles $a$. If the temperature $T$ can be kept
much less than these gaps $\Delta_a$ and inter-quasiparticle
distances $x$ kept much greater than the tunneling correlation lengths $\xi_a$, then the corresponding error rates will vanish as $e^{-{\Delta_a}/T}$ and $e^{-x/\xi_a}$ and, hence, be negligible.

The smallest gap for charged quasiparticles is usually assumed to correspond to the minimally charged excitations of a state~\footnote{This is generally a good assumption, given the Coulombic energy cost of forming quasiparticles with larger charge.}. For the MR and $\overline{\text{Pf}}$ states, the minimal charge $\pm e/4$ quasiparticles also carry non-Abelian Ising topological charge $\sigma$.
It is natural to interpret the gap corresponding to the temperature dependence of the longitudinal
resistance, $\rho_{xx}\sim e^{-\Delta_{\rm trans}/2T}$,
as the energy gap
$\Delta^{\rm qh}_\sigma + \Delta^{\rm qp}_\sigma$ for a charge $\pm e/4$
quasihole-quasiparticle pair,
which is thereby deduced from experiments to be
$\Delta^{\rm qh}_\sigma + \Delta^{\rm qp}_\sigma
\equiv \Delta_{\rm trans} \approx 0.5$K
in the highest-mobility samples~\cite{Choi08}.
Numerical studies of small numbers of electrons
interacting through Coulomb interactions in the second
Landau level find $\Delta^{\rm qh}_\sigma + \Delta^{\rm qp}_\sigma\approx 0.025-0.029\frac{e^2}{\varepsilon\ell_0}$ (which is
$3.2-3.7\,$K at $6.5$T)~\cite{Morf02,Feiguin09}.

However, bulk electrical transport is not sensitive to
the energy gap of electrically neutral excitations, such as the neutral fermion that carries Ising topological charge $\psi$ in the MR and $\overline{\text{Pf}}$ states. Consequently, $\Delta_\psi$ has not been measured
(though it could, in principle, be determined from
thermal transport measurements or, as we discuss
below, from interferometry measurements in mesoscopic
devices). $\Delta_\psi$ has previously not been theoretically
calculated, either.

The MR and $\overline{\text{Pf}}$ states are
the quantum Hall analogues of spin-polarized
$p_x+ip_y$ superconductors~\cite{Moore91,Greiter92,Read00}.
Charge $e/4$ quasiparticles $\sigma$ correspond to flux
$hc/2e$ vortices; neutral fermions $\psi$ correspond to
Bogoliubov-de~Gennes quasiparticles in the superconductor.
In most superconductors, these two gaps have completely different
scales and are not considered on the same footing.
However, in the $\nu=5/2$ state, there is only a single
energy scale ${e^2}/\varepsilon \ell_0$, so these gaps
can be comparable. Thus far, however, only $\Delta_\sigma$
has been computed. In this paper, we compute $\Delta_\psi$.
This is the appropriate quantity to use when comparing
the gap in the $\nu=5/2$ state to the gaps in other superconductors,
and when drawing lessons for non-phonon mechanisms of superconductivity from this state.

The neutral fermion gap is also a relevant quantity in
determining the effectiveness of topological protection
in the $\nu=5/2$ state.
The transfer of Ising $\psi$ charge between quasiparticles, e.g. through tunneling, alters the non-local state shared by
the two quasiparticles. It is, thus, responsible for splitting the degenerate non-local states and causing errors in the encoded information \cite{Bonderson09}.
Similarly, the neutral fermion gap directly determines the visibility of non-Abelian statistical signatures in interference experiments (see~\cite{Bishara09b}, and references therein), since tunneling of the neutral $\psi$ charge (between bulk quasiparticles and between bulk quasiparticles and the edge) suppresses interference terms. In this light, it is of paramount
importance to study this quantity. In this letter, we produce numerical estimates of the neutral fermion gap and correlation length for the $\nu=5/2$ non-Abelian quantum Hall state.

In order to model the $\nu=5/2$ state, we assume that
both spins of the lowest Landau level are filled and inert
and focus on the second Landau level, which has $\nu=1/2$.
Our calculation neglects finite layer-thickness~\cite{Peterson08} and Landau-level mixing~\cite{Bishara09a,Rezayi09,Wojs10},
which certainly play a role in real devices. A more realistic
calculation, including these effects, will be discussed elsewhere.
Here, we focus on the simplified situation of an infinitely-thin
two-dimensional layer in a very high magnetic field and study
small systems (${N_e}\leq 15$ electrons) by exact diagonalization
and larger systems ($13\leq {N_e}\leq 26$ electrons)
by the density-matrix renormalization
group (DMRG), as in~\cite{Feiguin08,Feiguin09,Shibata01}.

On the sphere, the MR ground-state occurs
at $N_{\phi} = 2 {N_e} -3$,
where the number of electrons ${N_e}$
should be taken to be even. For the following discussion, it will
be helpful to introduce the following terminology. For a given,
fixed arbitrary electron number ${N_e}$ and flux $N_\phi$, we will
call the state of lowest energy the ``lowest energy state.''
If ${N_e}$ is even and $N_{\phi} = 2 {N_e} -3$
we will call the lowest energy state
the MR
{\it ground state}. The reason for this
distinction is that for some values of ${N_e}$, $N_\phi$, the
lowest energy state should be understood as a state
with a quasiparticle excitation. We denote
the lowest energy for a system of ${N_e}$ electrons and $N_{\phi}$ fluxes
by $E\left( N_{\phi} , {N_e} \right)$.

In order to compute the energy gap for an electrically-neutral
quasiparticle, we need to compare
the usual ground-state to a configuration that forces the system to have a neutral excitation of non-trivial topological charge.
In an Ising-type system,
the lowest energy state on the sphere with
${N_e}$ odd electrons must have non-trivial quasiparticles whose total topological charge is $\psi$ \cite{torus}.
The two simplest possibilities are that such a state either has
a neutral $\psi$ quasiparticle, or a charge $e/4$ $\sigma$ quasihole
and $-e/4$ $\sigma$ quasiparticle pair that fuses into a $\psi$.
(If such a pair forms a bound state, it is equivalent to a neutral $\psi$ quasiparticle.) Which of these possibilities actually occurs depends on whether the energy $\Delta_\psi$ to create a neutral fermion $\psi$ is less than or greater than the energy $\Delta_{\sigma}^{\text{qh}} + \Delta_{\sigma}^{\text{qp}}$ to create the quasihole-quasiparticle pair.

Consequently, $E\left( N_{\phi}+2 , {N_e}+1 \right) -
E\left( N_{\phi} , {N_e} \right)$
is the energy $\mathcal{E}$ due to one electron plus/minus the energy of the non-trivial quasiparticle(s) for $N_e$ even/odd (either $\Delta_\psi$ or $\Delta_{\sigma}^{\text{qh}} + \Delta_{\sigma}^{\text{qp}}$).
To isolate the energy of these (collectively) neutral quasiparticle(s) with total topological charge $\psi$,
we define the neutral fermion gap
\begin{multline}
\label{eqn:Delta-psi-def}
\Delta_{F} \left( {N_e} \right) \equiv  \frac{\left(-1\right)^{N_e}}{2} \left[ E\left( N_{\phi}+2 , {N_e}+1 \right) \right.\\
\left. + E\left( N_{\phi}-2 , {N_e}-1 \right) -2 E\left( N_{\phi} , {N_e} \right) \right]
\end{multline}
for paired states. In the regime in which $E\left( N_{\phi} , {N_e} \right)$ scales linearly with ${N_e}$, the neutral fermion gap $\Delta_{F} \left( {N_e} \right)$ will be constant. It is instructive to contrast Eq.~\ref{eqn:Delta-psi-def}
with the expression for the charge gap,
$\Delta_{c} \left( {N_e} \right) = \frac{1}{2} \left[ E\left( N_{\phi}+1 , {N_e} \right) + E\left( N_{\phi}-1 , {N_e} \right) - 2 E\left( N_{\phi} , {N_e} \right) \right]$.
In Eq.~\ref{eqn:Delta-psi-def}, we compare the energies of systems
with the same charge-flux relation so that the net charge
of all excitations is zero while $\Delta_{c}$ compares the
energies of states with fluxes offset by one so that the net charge
of all excitations is $\nu e$.

\begin{figure}  [t!]
\includegraphics[width=3.5cm,angle=-90]{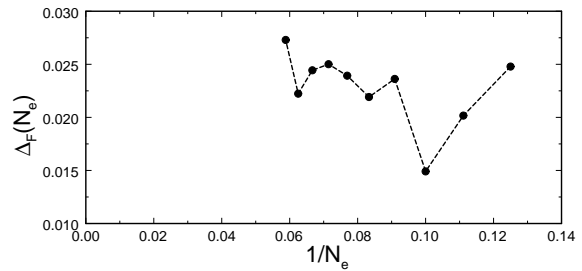}
\caption{The neutral fermion gap, as defined in
Eq.~(\ref{eqn:Delta-psi-def}), for the $\nu=5/2$
MR state as a function of inverse system size $1/{N_e}$.}
\label{fig:gap-neutral}
\end{figure}

For pure Coulomb interactions
in the second Landau level, we have computed the
ground state energies for even numbers of electrons
up to ${N_e}=26$ and the lowest state energies for
odd numbers of electrons up to ${N_e}=17$.
In a recent calculation, Lu \emph{et al.}~\cite{Lu10} have computed
these energies up to ${N_e}=18$ electrons by exact diagonalization;
our energies are in agreement with theirs.
In Fig.~\ref{fig:gap-neutral}, we show the values of the neutral fermion gap
${\Delta_F}({N_e})$, computed using Eq.~\ref{eqn:Delta-psi-def},
as a function of inverse system size $1/{N_e}$
for up to ${N_e}=17$ electrons.
As may be seen from Fig.~\ref{fig:gap-neutral},
the neutral fermion gap fluctuates considerably,
which is a sign of finite-size effects.
If we were to use a purely linear fit, then we would find
$\lim_{N_e \rightarrow \infty} \Delta_{F} (N_e) \approx 0.028$;
if we were to fit the gap to a constant, we would find
$\lim_{N_e \rightarrow \infty} \Delta_{F} (N_e) \approx 0.023$. However,
the errors in these fits, determined from the maximum fluctuation away
from the average, are large (though $\Delta_F$ is clearly non-zero).
Therefore, more care is needed in order to perform
an ${N_e} \rightarrow \infty$ extrapolation.

To this end, we note that if the system is gapped, then we can
write $E\left( N_{\phi} , {N_e} \right)$ in the form
\begin{equation}
\label{eqn:even-odd-energy}
E\left( N_{\phi} , {N_e} \right) =  \mathcal{E} {N_e} + E_{\text{even, odd}}
+ O(e^{-a\sqrt{{N_e}}})
\end{equation}
for $N_e$ even or odd, respectively.
The leading terms are the same for even and odd $N_e$
because the energy per particle $\mathcal{E}$ must be the same in the
thermodynamic limit. The constant terms $E_{\text{even,odd}}$ are due to the internal order of the phase and the genus of the system \cite{torus},
as well as the energy cost of the (collectively) neutral quasiparticle(s) for $N_e$ odd. Corrections to these first two terms are exponentially
small in the linear size of the system ($\sim\sqrt{{N_e}}$) since the system has a gap; here, $a$ is a
constant inversely proportional to the correlation length.

\begin{figure}  [t!]
\includegraphics[width=4.5cm,angle=-90]{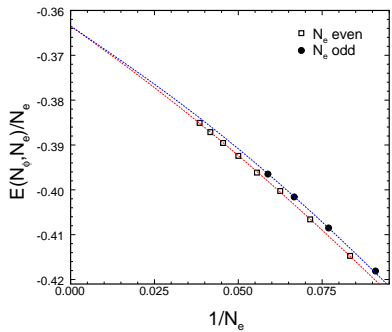}
\caption{The ${N_e} \rightarrow \infty$ extrapolation of $E\left( N_{\phi} , {N_e} \right)/{{N_e}}$ corresponding to the $\nu=5/2$ MR state for ${N_e}$ even (squares) and ${N_e}$ odd (dots) using Eq.~\ref{eqn:even-odd-energy},
from which we find $\Delta_{\psi} \approx 0.027$.}
\label{fig:energies}
\end{figure}

Substituting Eq.~\ref{eqn:even-odd-energy} into Eq.~\ref{eqn:Delta-psi-def},
we find
\begin{equation}
\Delta_{F} \left( {N_e} \right) = E_{\text{odd}} - E_{\text{even}} + O(e^{-a\sqrt{{N_e}}})
\end{equation}
and thus $\lim_{N_e \rightarrow \infty} \Delta_{F} (N_e)  = E_{\text{odd}} - E_{\text{even}}$, further justifying our definition of the neutral fermion gap. We can, however, use Eq.~\ref{eqn:even-odd-energy} to extract $E_{\text{odd}} - E_{\text{even}}$ more directly by simply fitting the numerical data with functions of this form, and it allows us to exploit
the larger system sizes for which we have computed
the ground state energies for even ${N_e}$. In Fig.~\ref{fig:energies},
we plot $E\left( N_{\phi} , {N_e} \right)/{{N_e}}$ vs. $1/{{N_e}}$, and fitting to Eq.~\ref{eqn:even-odd-energy} (divided by $N_e$)
but replacing, for simplicity, the $O(e^{-a\sqrt{N_e}})$ term by a single
term $c e^{-a\sqrt{N_e}}$,
we find $\mathcal{E} = -0.3634$, $E_{\text{even}} = -0.5381$, and $E_{\text{odd}} = -0.5114$. For $N_e$ even, we find $c = -0.7876$ and $a = 0.6675$, while for $N_e$ odd we find $c = -1.4700$ and $a = 0.8287$.
Thus, we can reliably extract the thermodynamic limit of the neutral
fermion gap by taking the difference between the $1/N_e$
terms in the expressions for $E\left( N_{\phi} , {N_e} \right)/{{N_e}}$.
We find $E_{\text{odd}} - E_{\text{even}} \approx 0.027$
(in units of $e^2/\varepsilon \ell_0$).

One advantage of using this method of extracting the neutral fermion gap is that
it is easier to diagnose potential difficulties with
the ${N_e} \rightarrow \infty$ extrapolation.
For instance, one potential pitfall is aliasing.
If one of the systems studied is actually in the ground state
of a different phase, then $E\left( N_{\phi} , {N_e} \right)/{{N_e}}$
would not sit on the expected (nearly-linear) curve. As may be seen from
the figure, the data points deviate negligibly from the fitting curves, so
this is not the case for the system sizes we study. At any rate,
the most serious potential aliases occur at ${N_e}<10$,
which we do not consider for the extrapolation in Fig.~\ref{fig:energies}

Having computed
$\lim_{N_e \rightarrow \infty} \Delta_{F} (N_e) \approx 0.027$ for the (thermodynamic limit of the) neutral fermion gap, we now address the
nature of the associated quasiparticle.
The statement that the $\nu=5/2$ state is an Ising-type topological state merely guarantees that
$\psi$ is an allowed value for the
topological charge in any region bounded by a closed
curve. It does not guarantee that there is actually
an energetically stable quasiparticle that has this value of topological charge.
The mismatch between allowed topological charges
and stable quasiparticle species is
is a feature of all topological states. For instance,
in the $\nu=1/3$ Laughlin state \cite{Laughlin83},
the charge $2e/3$ quasihole carries an allowed
value of topological charge, but it is not an energetically stable excitation
(for Coulomb interactions); if we attempt to create one,
it will decay into two charge $e/3$ quasiholes.
Similarly, we must consider the possibility that a neutral $\psi$ quasiparticle
will simply decay into a charge $\pm e/4$ $\sigma$ quasihole-quasiparticle
pair that fuses into the $\psi$ channel, i.e. that $\Delta_\psi > \Delta^{\rm qp}_\sigma + \Delta^{\rm qh}_\sigma$.
In this case, $\lim_{N_e \rightarrow \infty} \Delta_{F} (N_e) = E_{\text{odd}} - E_{\text{even}}$ would be identified with $\Delta^{\rm qp}_\sigma + \Delta^{\rm qh}_\sigma$ and provides a lower bound for $\Delta_\psi$.
However, since we find $E_{\text{odd}} - E_{\text{even}} \approx 0.027$
and previous studies~\cite{Feiguin08} obtained
$\Delta^{\rm qp}_\sigma + \Delta^{\rm qh}_\sigma \approx 0.029$,
we tentatively conclude that the neutral fermion $\psi$
is stable and has
\begin{equation}
\Delta_\psi = E_{\text{odd}} - E_{\text{even}} \approx 0.027
.
\end{equation}
Stronger evidence supporting this interpretation
comes from the good fit of our data to the $N_e$ odd case
of Eq.~\ref{eqn:even-odd-energy}.
If the $\psi$ quasiparticle were unstable, there would be a $-1/32 \sqrt{{N_e}}$
term in the odd electron number energy, resulting from the Coulomb interaction
energy between the $\pm e/4$ charges \cite{Morf02}.

For purposes of comparison, we note that a similar
computation of the neutral fermion gap for the $\nu=1/3$
Laughlin state \cite{Laughlin83} would give the value
zero because the even- and odd-electron number ground
state energies lie on the same line~\cite{Feiguin08}; since
it is not a paired state, there is no qualitative difference between even
and odd electron numbers.
On the other hand, the $\overline{\text{Pf}}$ state, the $(3,3,1)$ state~\cite{Halperin83}
and the Bonderson-Slingerland (BS) states~\cite{Bonderson08}
have neutral fermionic excitations whose gaps can be computed
by the method explained in this paper.
In the absence of Landau-level mixing, $\Delta_\psi$ is
expected to be precisely the same for the $\overline{\text{Pf}}$ state
as it is for the MR state; preliminary calculations are consistent
with this, as we report elsewhere \cite{Feiguin-forthcoming}. 
In the case of the $k \geq 3$ Read-Rezayi states~\cite{Read99},
there are neutral excitations that are
non-Abelian and, therefore, cannot be obtained
by simply altering the electron number and flux.

\begin{figure} [t!]
\includegraphics[width=3cm,angle=-90]{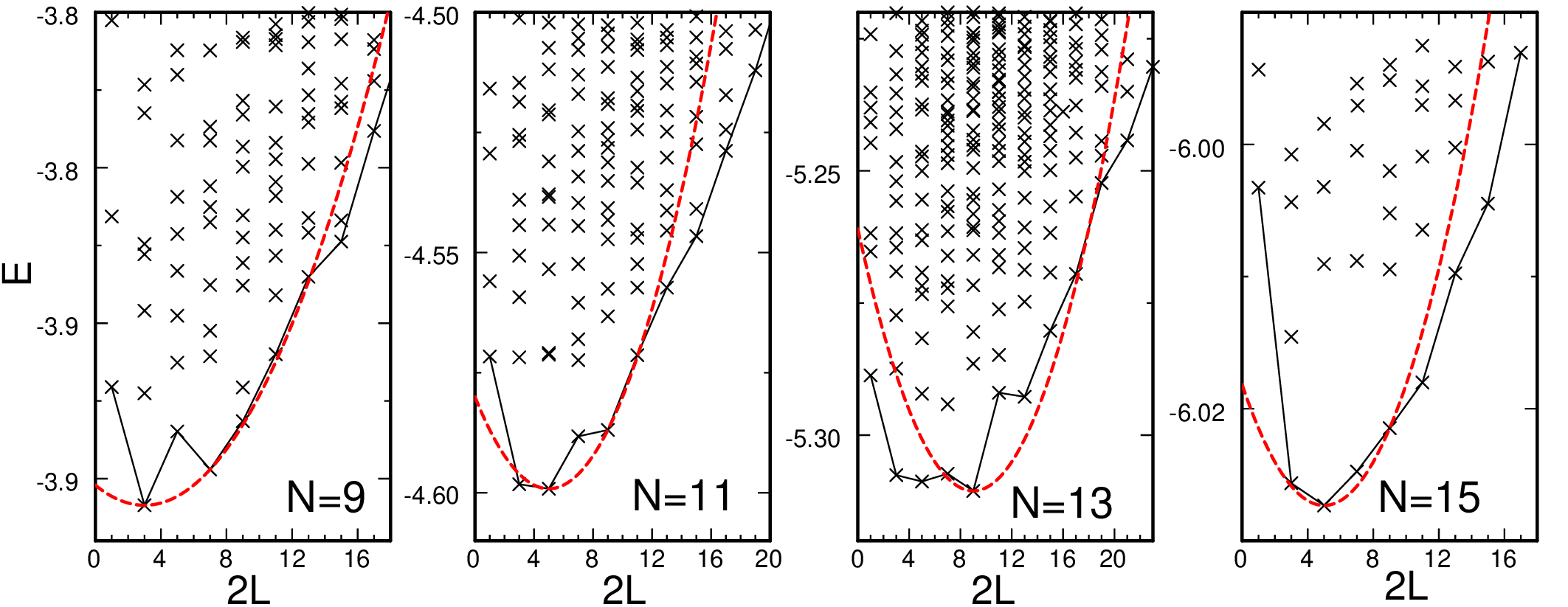}
\includegraphics[width=2.7cm,angle=-90]{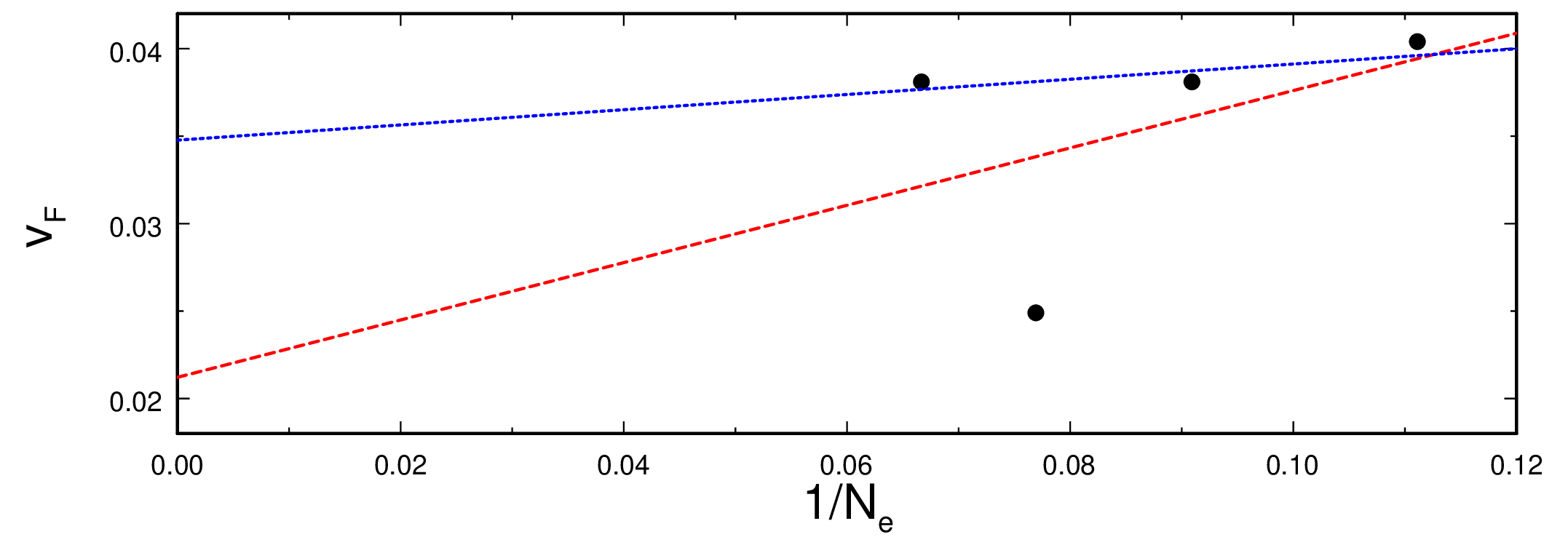}
\caption{The low-energy spectrum of the $\nu=5/2$ MR state for
$9,11,13,15$ electrons with parabolic fits of the lowest-lying
states to Eq.~\ref{E-vs-L}, from which we extract the
effective velocities plotted in the lower panel.}
\label{fig:low-energy-spec}
\end{figure}

Although the neutral fermion gap has not
been previously calculated, a related quantity
has recently been calculated, namely the
splitting between the two degenerate states that occur for four $e/4$ $\sigma$ quasiparticles~\cite{Baraban09}. This splitting, $\Delta E(r)$,
decays with distance $r$ between the $\sigma$ quasiparticles
as $\Delta E(r) \sim f(r)\, e^{-r/\xi}$
for large $r$. Here, $f(r)$ is an oscillatory function
and $\xi$ is the characteristic length scale for the decay.
If we interpret this splitting as the energy associated
with inter-quasiparticle tunneling of neutral fermions, then we expect
$\xi=\xi_{\psi} = v/\Delta_\psi$,
where $v$ is the velocity of a neutral fermion.
If the MR state is interpreted as a
paired state with small gap, then $v$ would be the Fermi
velocity $v_F$ of the underlying Fermi-liquid-like
metallic state. In such a case, the Fermi velocity could
be deduced by studying the spectrum of a single neutral fermion
as follows. For odd ${N_e}$, the energy spectrum
will not have a gap above the lowest energy state
(in the thermodynamic limit) since there will
be one unpaired neutral fermion above the Fermi energy,
and this fermion can be excited to any other state above
the Fermi energy. In a BCS mean-field theory, the energy
spectrum for odd $N_e$ will be bounded below by the curve
${E_L} = \sqrt{\epsilon_L^2 + \Delta_\psi^2} + E_{\rm gs}
\approx  \frac{1}{2\Delta_F}\epsilon_{L}^2 + {\Delta_F} + E_{\rm gs}$,
where $E_{\rm gs}$ is the ground state energy for ${N_e}-1$
electrons, $\epsilon_L$ is a single-particle energy relative
to the Fermi energy for a state with angular momentum $L$.
We take $\epsilon_L = \frac{v_F}{N_\phi \ell_0} [L(L+1) - L_0(L_0 + 1)]$,
where $L_0$ is the highest occupied angular momentum
orbital. Thus, for $L\approx L_0$, the excitation energies are expected
to be quadratic in $L-L_0$:
\begin{equation}
\label{E-vs-L}
E_L \approx  \frac{1}{2\Delta_F}
\left(\frac{v_F (2L_0 +1)}{N_\phi \ell_0}\right)^2 (L-L_0)^2
\,+ \,\text{const.}
\end{equation}
As may be seen in Fig.~\ref{fig:low-energy-spec},
the lowest excitation energies for ${N_e}=9,11,13,15$
appear to follow a parabola. A linear extrapolation
of the $v_F$ values obtained from these spectra
according to Eq.~\ref{E-vs-L}
gives $v_F \approx 0.021\, {e^2}/\varepsilon$, which leads to
$\xi_\psi \approx 0.8\,\ell_0$. However, the parabolic fit is quite
poor for $N=13$; the other three system sizes are consistent
with $v_F \approx 0.035\, {e^2}/\varepsilon$, or $\xi_\psi \approx 1.3\,\ell_0$.
We note, for comparison, that Baraban {\it et al.} \cite{Baraban09}
find a length scale $\xi\approx 2.3 \ell_0$,
although their calculation is for much larger system sizes
and for trial wavefunctions, rather than the Coulomb ground state.

Our results imply that a quantum computer
based on the $\nu=5/2$ fractional quantum Hall state should
be operated at temperatures much lower than $\Delta_\psi$,
which is $\approx 3.4$K for a magnetic field $B=6.5$T.
This implies that, at $35$mK, the rate at which
phase errors can be expected for a topological qubit
is $\sim e^{-\Delta_\psi /T}\approx 10^{-44}$
if the computational anyons are further than $\xi_\psi\approx 130$\AA
{} from each other and the edge.
In the experiments of Willett~{\it et al.}~\cite{Willett09},
the inter-quasiparticle distances are probably comparable to $\xi_\psi$; this
implies that the error rate may be large and
there is probably significant splitting between
the $2^{n-1}$ states expected for $2n$ quasiparticles~\cite{Nayak96c}.
By measuring the time over which the signal through
an interferometer remains stable, it should be possible
to measure the error rate and, thereby, $\Delta_\psi$.
In addition, bulk thermal transport may be dominated
by thermally-excited neutral fermions. Although charge $e/4$
quasiparticles may have a smaller energy gap (approximately
half that for a $\psi$), they will be much more strongly
localized by disorder than neutral fermions.

Finally, we note that $\Delta_\psi \approx 0.027 \,
\frac{e^2}{\varepsilon \ell_0}$ is small compared to
the Coulomb energy. Since one might argue that
the gap is small because of the proximity to competing phases,
such as the striped phase~\cite{Rezayi00},
we consider the neutral fermion gap for
a Hamiltonian in which the only interaction
is the (repulsive) three-body interaction for which the MR wavefunctions
are the exact ground states~\cite{Greiter92,Rezayi00}. For this Hamiltonian,
the ground state energy is precisely zero for
${N_e}$ even, so $E_{\text{even}}=0$. Thus, we must only compute
the ground state energies for ${N_e}$ odd. A simple linear extrapolation
of these energies
gives $\lim_{N_e \rightarrow \infty}\Delta_F = E_{\text{odd}} \approx 0.45$,
if the coefficient of the three-body interaction is $1$.
Thus, there is nothing wrong in principle with the na\"ive
idea that the superconducting gap can be comparable to
the Coulomb energy scale for an electronic pairing mechanism,
so long as there are no nearby competing phases to suppress it.

\acknowledgements
We would like to thank M. Hastings, M. Peterson,
and E. Rezayi for very helpful discussions and the Aspen Center for
Physics for hospitality. A.F. is supported by the NSF grant
DMR-0955707 and C.N. by the DARPA-QuEST program.


\begin{thebibliography}{10}

\bibitem{Willett87}
R. Willett {\it et~al.}, Phys. Rev. Lett. {\bf 59},  1776  (1987).

\bibitem{Pan99b}
W. Pan {\it et~al.}, Phys. Rev. Lett. {\bf 83},  3530  (1999).

\bibitem{Eisenstein02}
J.~P. Eisenstein {\it et al.}, Phys. Rev.
  Lett. {\bf 88},  076801  (2002).

\bibitem{Kitaev97}
A.~Y. Kitaev, Ann. Phys. (N.Y.) {\bf 303},  2  (2003).

\bibitem{Freedman98}
M.~H. Freedman, Proc. Natl. Acad. Sci. U.S.A. {\bf 95},  98  (1998).

\bibitem{Nayak08}
C. Nayak {\it et~al.}, Rev. Mod. Phys. {\bf 80},  1083  (2008).

\bibitem{Morf98}
R.~H. Morf, Phys. Rev. Lett. {\bf 80},  1505  (1998).

\bibitem{Rezayi00}
E.~H. Rezayi and F.~D.~M. Haldane, Phys. Rev. Lett. {\bf 84},  4685  (2000).

\bibitem{Feiguin09}
A.~E. Feiguin {\it et~al.}, Phys. Rev. B {\bf 79},  115322  (2009).

\bibitem{Bishara09a}
W. Bishara and C. Nayak, Phys. Rev. B {\bf 80},  121302  (2009).

\bibitem{Rezayi09}
E.~H. Rezayi and S.~H. Simon, arXiv.org:0912.0109.

\bibitem{Feiguin08}
A.~E. Feiguin {\it et al.}, Phys. Rev. Lett. {\bf
  100},  166803  (2008).

\bibitem{Peterson08}
M.~R. Peterson {\it et al.}, Phys. Rev. Lett. {\bf 101},
  016807  (2008).

\bibitem{Radu08}
I. Radu {\it et~al.}, Science {\bf 320},  899  (2008).

\bibitem{Dolev08}
M. Dolev {\it et~al.}, Nature {\bf 452},  829  (2008).

\bibitem{Willett09}
R.~L. Willett {\it et al.}, Proc. Nat.
  Acad. Sci. (USA) {\bf 106},  8853  (2009).

\bibitem{Moore91}
G. Moore and N. Read, Nucl. Phys. B {\bf 360},  362  (1991).

\bibitem{Greiter92}
M. Greiter {\it et al.}, Nucl. Phys. B {\bf 374},  567  (1992).

\bibitem{Lee07}
S.-S. Lee {\it et al.}, Phys. Rev. Lett. {\bf 99},
  236807  (2007).

\bibitem{Levin07}
M. Levin {\it et al.}, Phys. Rev. Lett. {\bf 99},  236806
  (2007).

\bibitem{Choi08}
H.~C. Choi {\it et~al.}, Phys. Rev. B {\bf 77},  081301  (2008).

\bibitem{Morf02}
R.~H. Morf {\it et al.}, Phys. Rev. B {\bf 66},
  075408  (2002).

\bibitem{Read00}
N. Read and D. Green, Phys. Rev. B {\bf 61},  10267  (2000).

\bibitem{Bonderson09}
P. {Bonderson}, Phys. Rev. Lett. {\bf 103},  110403  (2009).

\bibitem{Bishara09b}
W. Bishara {\it et al.}, Phys. Rev. B {\bf 80}, 155303 (2009).

\bibitem{Wojs10}
A. W\'ojs {\it et al.}, Phys.
  Rev. Lett. {\bf 105},  096802  (2010).

\bibitem{Shibata01}
N. Shibata and D. Yoshioka, Phys. Rev. Lett. {\bf 86}, 5755 (2001);
J.~Phys.~Soc.~Jpn. {\bf 72}, 664 (2003).

\bibitem{Lu10}
H. {Lu} {\it et al.}, arXiv:1008.1587.

\bibitem{Laughlin83}
R.~B. Laughlin, \prl {\bf 50}, 1395 (1983).

\bibitem{Halperin83}
B.~I. Halperin, Helv. Phys. Acta {\bf 56},  75  (1983).

\bibitem{Bonderson08}
P. Bonderson and J.~K. Slingerland, Phys. Rev. B {\bf 78},  125323  (2008).

\bibitem{Feiguin-forthcoming}
A. Feiguin {\it et al.}, in preparation.

\bibitem{Read99}
N. Read and E. Rezayi, Phys. Rev. B {\bf 59},  8084  (1999).

\bibitem{Baraban09}
M. Baraban {\it et al.}, Phys. Rev. Lett. {\bf
  103},  076801  (2009).

\bibitem{Nayak96c}
C. Nayak and F. Wilczek, Nucl. Phys. B {\bf 479},  529  (1996).

\bibitem{torus}
On the torus, an odd number of electrons
can be accommodated without creating quasiparticles.
This agrees with the fact that the curvature is zero,
so there are no constant terms in Eq.~\ref{eqn:even-odd-energy}.

\end{thebibliography}

\end{document}